\documentclass[twocolumn,prl,showpacs,superscriptaddress,amsmath,amssymb]{revtex4-1}
\usepackage[utf8]{inputenc}
\usepackage{graphicx,color} 
\usepackage{latexsym}
\usepackage{mathtools}

\renewcommand{\vec}[1]{\boldsymbol{#1}}

\newcommand{\caX}{{\mathcal X}}

\newcommand{\diff}{\text{d}}

\newcommand{\mean}[1]{{\left< #1 \right>}}

\begin{document}

\title{Exact symmetries in the velocity fluctuations of a hot Brownian swimmer}

\newcommand\uleip{\affiliation{Institut f\"ur Theoretische Physik, Universit\"at Leipzig,  Postfach 100 920, D-04009 Leipzig, Germany}}
\newcommand\uleipexp{\affiliation{Molecular Nanophotonics Group, Institute of Experimental Physics I, University of Leipzig, 04103 Leipzig, Germany}}
\newcommand\MPI{\affiliation{Max Planck Institute for Mathematics in the Sciences, Inselstr. 22, 04103 Leipzig, Germany}}

\author{Gianmaria Falasco}\MPI
\author{Richard Pfaller}\uleip
\author{Andreas P. Bregulla}\uleipexp
\author{Frank Cichos}\uleipexp
\author{Klaus Kroy}\uleip
\date{\today}


\begin{abstract}

Symmetries constrain dynamics. We test this fundamental physical
principle, experimentally and by molecular dynamics simulations, for a
hot Janus swimmer operating far from thermal equilibrium.  Our results
establish scalar and vectorial steady-state fluctuation theorems and a
thermodynamic uncertainty relation that link the fluctuating particle
current to its entropy production at an effective temperature. A
Markovian minimal model elucidates the underlying non-equilbrium physics.

\end{abstract}

\maketitle

Throughout the 19th century, Brownian motion was studied intensively
but remained a confusing and enigmatic phenomenon.  Even excessive
experimental data for the velocity of colloidal particles did not seem
to allow for a conclusive characterization of the motion, until
interpreted as mean-square spatial displacements within a consistent
theory of Brownian fluctuations \cite{ein05}. The main point
is that the mean-square displacement is robust with respect
to the chosen time discretization of the trajectory, while the
velocity is not. Nevertheless, the infamous velocity of Brownian
particles has recently gained renewed conceptual and practical
interest. This is partly due to improved technical abilities to assess
its regular short-time limit \cite{fra11,khe14}, and partly
to a surge of applications involving self-propelled colloidal particles
\cite{mar13}. The directed autonomous motion of such ``Brownian swimmers''
can be harnessed for performing mechanical work and other potentially
useful tasks \cite{heb10}, but its persistence is limited by Brownian
fluctuations. In the following,
we demonstrate that these fluctuations are not completely chaotic but
encode a characteristic fingerprint of the underlying space-time
symmetry, even far from thermal equilibrium.

We specifically consider a so-called hot Brownian swimmer, in theory,
experiment, and computer simulations.  The swimmer is designed as a
Janus sphere made of two hemispheres with unequal thermal and
solvation properties, which excites phoretic surface flows upon
heating. A variety of such phoretic self-propulsion mechanisms is
commonly employed in the design of artificial microswimmers.  They all
rely on the creation of asymmetric gradients of a thermodynamic field
(e.g.\ concentration of a solute \cite{gol07b}, temperature
\cite{sano10}) in the solvent, which induces a systematic drift
through classical interfacial phoretic processes
\cite{anderson}. Besides their biomimetic and practical relevance for
promising applications in nano-science, self-propelled particles are
of fundamental interest as paradigmatic non-equilibrium systems. Their
energy input is not due to fluxes at the boundary, as often the case
in macroscopically induced non-equilibrium states, but localized on
the particle scale.  And their self-propulsion is not due to a balance
of external body forces with Stokes friction, as assumed in popular
``dry-swimmer'' models \cite{mar13}.  Instead, they are characterized
by (hydrodynamic) long-range interactions very different from those
found in driven colloids.  In other words, microswimmers are not simply
ordinary driven colloids in disguise.  As a result, their collective
behavior displays peculiar features, such as phases and phase
transitions absent in sheared and sedimenting colloids, say
\cite{gol12,theurkauff12}. But clear non-equilibrium signatures are
detected already on the single-particle level. Examples include
non-Gaussian fluctuations \cite{loewen13}, negative mobility for
confined swimmers \cite{gho14}, and hot Brownian motion
\cite{rin10} for non-isothermal swimmers. Despite the possibility to
experimentally track and manipulate single particle trajectories
\cite{cichos14}, a less explored direction of research is the study of
fluctuations of (thermodynamic) path observables, defined along
trajectories, as in stochastic energetics \cite{sek10,sei12}. In
particular, one may wonder whether fluctuation theorems are valid for
self-propelled particles.

Fluctuation theorems are symmetry relations holding for a wide class
of non-equilibrium systems
\cite{eva93,gal95,kur98,leb99,mae99,sei05}. They quantify the
irreversibility of non-equilibrium processes by their total entropy
production $S$, saying that the  probability $P(S)$ for positive $S$
is exponentially larger than for negative $S$:
\begin{align}\label{ft}
P(S)=P(-S) e^{S/ k_B}.
\end{align}
Such relations ultimately originate from the time-reversal invariance
of the microscopic dynamics that is broken at the ensemble level by a
non-conservative driving. More recently, additional fluctuation
theorems were identified, which rely on the breaking of \emph{spatial}
symmetries of the underlying microscopic dynamics \cite{hu11, hur14,kum15}.
They relate the probabilities of vectorial observables pointing in
different directions, such as the isometric currents $\vec J$ and
$\vec J'$ (currents of equal strength in different directions) excited
by a homogeneous driving force $\vec \epsilon$ in an isotropic system:
\begin{align}\label{fts}
P(\vec J)=P(\vec J') e^{\vec \epsilon \cdot (\vec J - \vec J')} \,.
\end{align}


For our hot Brownian swimmers (and other self-propelled particles),
the validity of Eqs.~\eqref{ft} and~\eqref{fts} is \emph{a priori} in
doubt. First, the phoretic mechanism responsible for the entropy
production is not an external deterministic driving, as usually
assumed to derive Eqs.~\eqref{ft}~and~\eqref{fts}. In addition, due to
the presence of strong and long-ranged thermodynamic gradients, the
solvent is not an equilibrium bath. Hence, the thermal noise need not
be Gaussian as invoked in proofs of fluctuation relations for noisy
dynamics. In case of a particle generating a temperature gradient, the
bath noise does not even possess a definite temperature but is
generally characterized by a nontrivial noise temperature spectrum
arising from hydrodynamic memory \cite{jol11, fal14}. Finally,
Eq.~\eqref{fts} has been proved so far only for models that do not
include inertia \cite{esp15}. Despite all that, we now verify the
fluctuation relations \eqref{ft} and \eqref{fts} relating the entropy
production and the particle current for a hot Brownian swimmer both
experimentally and numerically. A minimal analytical model helps to
rationalize our findings.

The experimental system consists of a polystyrene bead of
$500~\text{nm}$ radius, half-coated with a $50~\text{nm}$ thick gold
layer, in aqueous solution. It is narrowly
confined between two glass plates coated with a non-ionic surfactant
(Pluronic) to prevent the particle from sticking.  The sample
is illuminated through a dark field condenser, and the
scattered light is collected and imaged with a CCD-camera. The
particle's center of mass position $\vec r$ and orientation $\vec n$
(a unit vector along the symmetry axis from hot to cold) are recorded
at an inverse frame rate of $5~\text{ms}$.  A $532~\text{nm}$ laser with
incident intensity of $0.05~\text{mW}/\mu\text{m}^2$ continuously heats the
particle.  The piezo-position employed for the spatial positioning of
the sample is adjusted every $100$ frames to keep the particle in the
center of the Gaussian beam.  The tangential surface gradient of the
local temperature translates into a thermophoretic propulsion velocity
$v_{\text{p}} \vec n$, stemming from the unbalanced particle-fluid
interactions \cite{anderson} plus thermoosmotic contributions from the
nearby glass covers \cite{bre16}.

\begin{figure}
\centering
\includegraphics[width=\columnwidth]{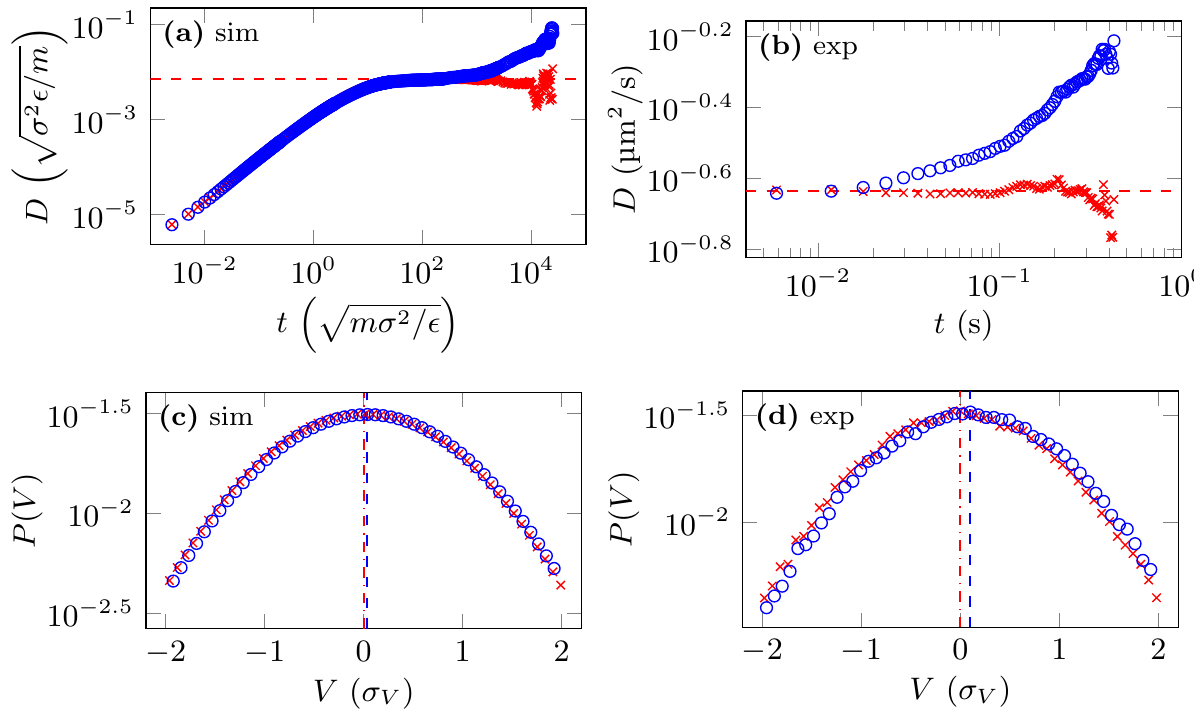}
\caption{\emph{Upper panel:} longitudinal (\color{blue}$\circ$\color{black}) and transverse ({\footnotesize{\color{red}$\times$\color{black}}})
  time-dependent diffusivities $D_\|$ and $D_\perp$  in the
  particle frame, as measured
  in simulation (a) and experiment (b). Their common
  plateau gives the  diffusion constant $D$, and the late-time slope of
  $D_\|(t)$ the propulsion speed $v_\text{p}$. \emph{Lower panel:}
  histograms of the corresponding
  short-time displacements. }
\label{fig:vd}
\end{figure}

A similar, but technically somewhat different realization of a hot
swimmer was implemented numerically. We performed molecular dynamics
simulations of $6\cdot 10^4$ Lennard--Jones (LJ) atoms of mass $m$. The
$255$ atoms belonging to the spherical swimmer were additionally
bonded by a FENE-potential \cite{chak11}. The standard LJ potential
$4\epsilon\bigl[(\sigma/r)^{12}-c_{i f}(\sigma/r)^6\bigr]$ was
modified by introducing an effective wetting parameter $c_{if}$ for
the interactions between the fluid ($f$) and the atoms composing the
two hemispheres ($i=a,\, b$) of the particle surface. They serve to
mimic the asymmetric solvent heating by the gold cap and polystyrene
body of the experimental particle, in an efficient way \cite{sch15}.
We used $c_{af}=2$ and $c_{bf}=1$ for all simulations reported in this
paper.  In a typical simulation run, we first equilibrate the system
in the Nos\'{e}-Hoover NVT ensemble at the prescribed temperature
$T=0.75 \epsilon/k_B$ and with an average number density
$\sim0.8\sigma^{-3}$ of fluid particles. Subsequently, we switch off the
global thermostat and only the atoms near the boundary of the
simulation box are kept at the ambient temperature, whereas the atoms
on one half of the particle surface are heated to a substantially
elevated temperature $T_p=1.5\epsilon/k_B$, by means of a standard
velocity rescaling algorithm.  After allowing the system to reach a
steady state, we record time traces of the position $\vec r$ and
velocity $\vec V$ of the particle's center of mass, as well as the
particle orientation $\vec n$. Differently from the experiment, the
system is spatially isotropic in three dimensions.

We start our analysis with the time-dependent diffusion coefficients
parallel and perpendicular to $\vec n$, 
\begin{align}
&D_{\parallel}(t)= \frac{1}{2} \frac{\diff}{\diff t}\mean{\left(\int_0^t \diff \tau \vec V(\tau) \cdot \vec n(\tau)\right)^2}, \\
&D_{\perp}(t)= \frac{1}{4} \frac{\diff}{\diff t}\mean{\left( \int_0^t \diff \tau \vec V(\tau) \cdot (\vec 1-\vec n\vec n(\tau))\right)^2},
\end{align}
where $\mean{\dots}$ denotes steady-state ensemble averaging.  As seen
from Fig.~\ref{fig:vd}a, the particle performs free diffusion in the
perpendicular direction on times beyond the timescale
$\tau_\text{St}$ ($\approx 10\sqrt{m\sigma^2/\epsilon}$ in the simulations) for
velocity damping, given by the particle mass over the Stokes
friction. This manifests itself as a plateau in $D_{\perp}(t)$, which
can be used to read off the long-time translational diffusion constant
as $D \equiv D_{\perp}( t \gg \tau_\text{St})$.  Along the particle
axis, the active drift $v_{\text{p}}=\mean{\vec n \cdot \vec V}$
(technically extracted from recorded trajectories by averaging over realizations
and time) is superimposed as a ballistic component, so that
$D_{\parallel}(t\gg \tau_\text{St}) \sim D + v_{\text{p}}^2
t$. Nevertheless, the mixing of parallel and perpendicular dynamics in
the lab frame randomizes the particle orientation $\vec n(t)$ at late
times, giving rise to an enhanced apparent overall diffusion
coefficient. Note that all mentioned diffusion and mobility
coefficients are non-equilibrium transport coefficients, since
(global) thermal equilibrium is broken.

The data in Figs.~\ref{fig:vd}c,d show that, despite the substantial
thermal gradients attained in the simulations, the particle velocities
parallel and perpendicular to $\vec n$ are essentially Gaussian
distributed. This demonstrates that the thermal agitation of the
Janus particle can be attributed to a Gaussian noise. It
also implies that, at least for the swimmer, the fluid can
effectively be described as locally in thermal equilibrium, which is indeed
a necessary assumption in standard theories of phoretic transport and
non-isothermal Langevin descriptions.

From the above observations, the following picture for the physics
underlying Eqs.~(\ref{ft}) and \eqref{fts} emerges. On average, the
particle will be propelled along the axis $\vec n$, or more generally,
such that $\vec n \cdot \dot{\vec r} >0$.  But on rare occasions, a
fluctuation can displace the particle against the phoretic drift, such
that $\vec n \cdot \dot{\vec r} <0$. The two dissimilar situations
correspond to energy dissipation to the fluid and energy extraction
from it, respectively. While the first conforms with the expected
thermodynamic behavior, the second represents an atypical transient
fluctuation.  Their relative rate is exactly quantified by
Eq.~\eqref{ft}.

To formalize this intuitive picture, we now propose a minimal model
for the swimmer dynamics. By restricting our analysis to times $t \gg
\tau_\text{St}$, the Brownian fluctuations are effectively diffusive,
and the particle momentum and all hydrodynamic modes can be taken as
fully relaxed. Long-time tails and any randomness in the swimming
speed $v_{\text{p}}$, which may arise from the fluctuating fluid
momentum and temperature, are discarded. On this level, the stochastic
motion of the hot Janus particle can be represented by the caricature
of two isotropic Markov processes for its position and orientation
vectors $\vec r(t)$ and $\vec n(t)$, with a superimposed
constant drift along $\vec n(t)$. The corresponding overdamped
Langevin equations read \footnote{To respect the Markovian
  approximation, the time derivatives have to be interpreted as
  discrete variations over finite time increments $\Delta t \gg
  \tau_\text{St}$.}
\begin{align}
&\dot{\vec r}= v_{\text{p}} \vec n + \sqrt{2D} \vec \xi_{\rm t}\, , \label{Langevin1} \\
&\dot{\vec n}= \sqrt{2D_{\rm r}}\vec \xi_{\rm r} \times \vec
  n\, , \label{Langevin2} 
\end{align}
where $\vec \xi_{\rm t}(t)$ and $\vec \xi_{\rm r}(t)$ represent independent
unbiased Gaussian white noise processes with unit variance.

\begin{figure}
\centering
\includegraphics[width=\columnwidth]{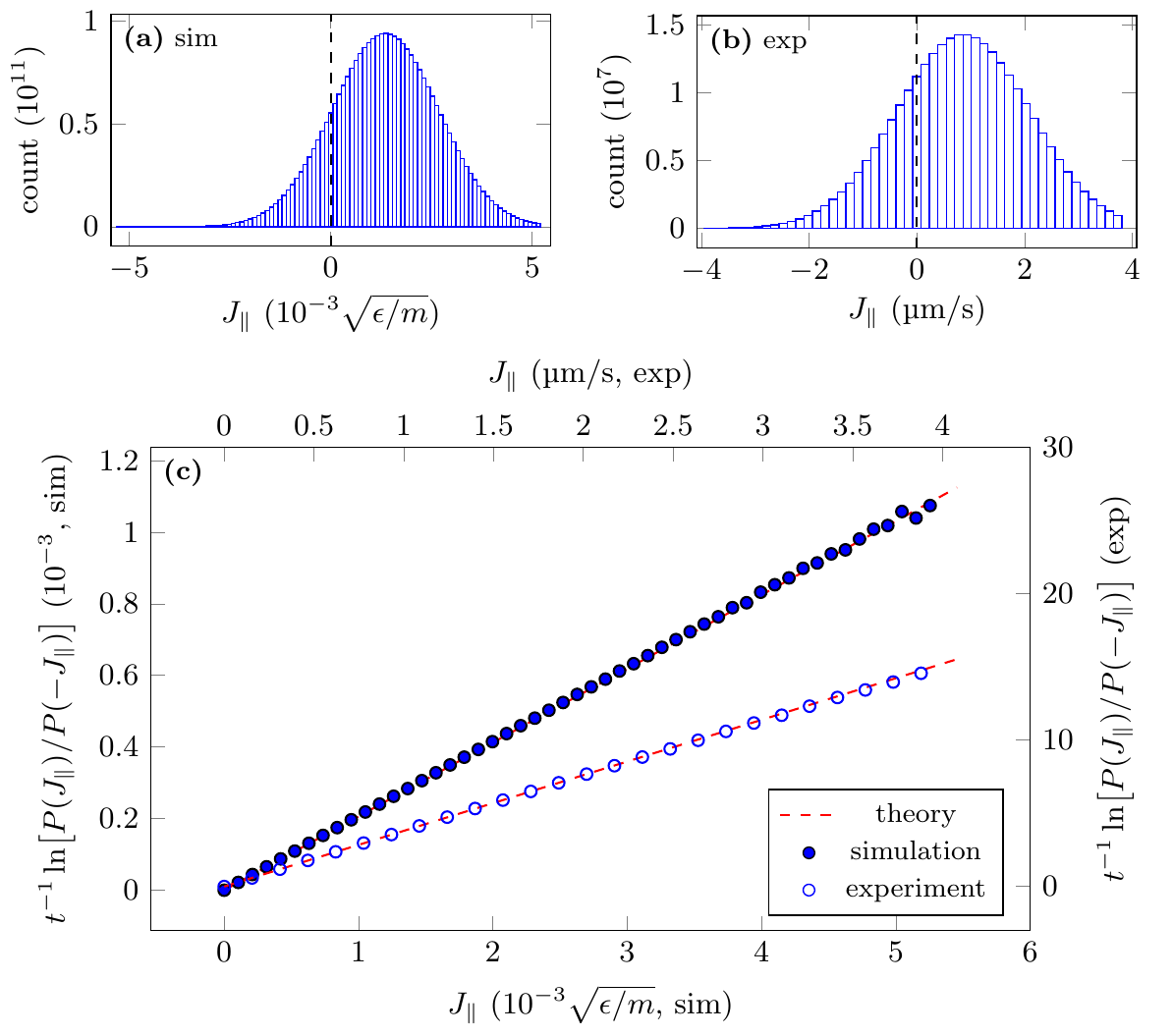}
\caption{\emph{Upper panel:} histograms of the longitudinal particle
  current $J_\|$, Eq.~(\ref{v_parallel}),  as measured in simulation (a) and
  experiment (b). \emph{Lower panel:} test of the fluctuation theorem,
  Eq.~(\ref{logratio}), using these histograms as proxies for $P(J_\|)$.}
\label{fig:scalarFR}
\end{figure}

We next consider the probability 
\begin{align}\label{P}
P[{\caX}] \propto \exp{\left(-\frac{1}{4D} \int_0^t \diff \tau\, [\dot{\vec
      r}(\tau)-v_{\text{p}} \vec n(\tau)]^2\right) } .
\end{align}
associated with a path ${\caX}\equiv \{\{\vec r(\tau), \vec n(\tau)\}:
0\leqslant \tau \leqslant t\}$ (the ordered set of positions and
orientations in the time interval $[0,t]$).  The path weight given by
Eq.~\eqref{Langevin2} is omitted, because it is inessential, as is the
initial configuration $\{\vec r(0), \vec n(0)\}$, because all allowed
configurations are equiprobable for an unconfined particle in the
steady state. The probability $P[\tilde{\caX}]$ to observe the same
event backwards in time is obtained from Eq.~\eqref{P} by the time
reversal transformation $\tau \to t-\tau$.  The two path weights are thus
related by
\begin{align}\label{ratio}
P[{\caX}] = P[\tilde{\caX}] \exp{\left(\frac{v_{\text{p}}}{D} \int_0^t
  \diff \tau \; \dot{\vec r}(\tau) \cdot  \vec n(\tau) \right) }\, ,
\end{align}
saying that a path ${\caX}$ resulting in a positive total
displacement along $\vec n$ is exponentially more probable than the
reversed path $\tilde{\caX}$ resulting in a negative
displacement. To show that this leads to a measurable asymmetry, we
define a time-averaged forward velocity
\footnote{This formal expression should be interpreted in a discrete
  sense, with the integrand approximated according to the midpoint
  (i.e. Stratonovich) rule $\dot{\vec r}(\tau) \cdot \vec n(\tau) \diff \tau
  \simeq [\vec r(s+\Delta s ) -\vec r(\tau)] \cdot [\vec n(s+\Delta
    s)+\vec n(\tau)]/2$.}
\begin{align}\label{v_parallel}
j_\|[{\caX}]  & \equiv  \frac 1 t \int_0^t  \diff \tau \; \dot{\vec r}(\tau) \cdot  \vec n(\tau).
\end{align}
The probability that $j_\|[\caX]$ attains a certain value, say $J_\|$,
follows by multiplying both sides of Eq.~\eqref{ratio} by
$\delta(j_\|-J_\|)$ and summing over all possible trajectories:
\begin{align}\label{Sum}
\sum_{{\caX}} P[{\caX}] 
 &\delta(j_\|- J_\|)=  \sum_{{\caX}} P[\tilde{\caX}]
e^{\frac{v_{\text{p}}}{D} t j_\| } \delta(j_\| -J_\|).
\end{align}
The left-hand side of Eq.~\eqref{Sum} is, by definition,
$P(J_\|)\equiv \textrm{Prob}(j_\|[\caX] = J_\|)$, while the right-hand side becomes
\begin{align*}
 \sum_{ \tilde{\caX}} P[\tilde{\caX}] e^{-\frac{v_{\text{p}}}{D}t j_\| } \delta( j_\| +J_\|)
  &=  P(-J_\|)  e^{\frac{v_{\text{p}}}{D} t J_\| } 
 \end{align*}
when we rewrite the sum over paths as sum over time-reversed
paths and flip the sign of $j_\|$ accordingly, since 
$j_\|[{\caX}] = - j_\|[\tilde{\caX}]$. We thus obtain the fluctuation relation
\begin{equation}\label{logratio}
\frac{1}{t}\ln \frac{P(J_\|)}{ P(-J_\|)}=\frac{v_{\text{p}}}{D} J_\|.
\end{equation}
It is valid for all observation times consistent with the Markov
condition $t \gg \tau_\text{St}$.  Differently from usual steady-state
fluctuation relations, boundary terms corresponding to the density of
states evaluated at times $0$ and $t$ are absent, being trivial
constants. This enables us to verify the theory at relatively short
times, which permits an efficient sampling of the negative tail of
$P(J_\|)$.  Corresponding histograms constructed from the numerical
and experimental data are shown in Fig.~\ref{fig:scalarFR}a,~b.  The
logarithmic ratios of statistically relevant opposite bins
conform nicely with Eq.~(\ref{logratio}), as shown in Fig.~\ref{fig:scalarFR}c.

%
%
%

The argument of the exponential in Eq.~\eqref{ratio}
may be cast into the explicit form of an entropy production by
combining Eq.~\eqref{ratio} with the generalized Sutherland--Einstein
relation, $D=k_B T_{\text{eff}}/\zeta$. This amounts to replacing
the non-isothermal solvent by a virtual isothermal bath at an effective temperature
$T_{\text{eff}}$ \cite{fal14}. Thereupon, Eq.~(\ref{ratio}) takes the
form of Eq.~\eqref{ft} with $P(S) \equiv \textrm{Prob}(s[\caX]=S)$
and
\begin{align}\label{S}
s[{\caX}] \equiv \frac{1}{T_{\text{eff}} }\int_0^t  \diff \tau \;
\dot{\vec r}(\tau) \cdot \vec n(\tau)  v_{\text{p}}\zeta =
\frac{ j_\| v_\text{p}\zeta}{T_\text{eff}}t 
\end{align}
the entropy produced by the ``thermophoretic force''
$\vec n v_{\text{p}}\zeta$ (the phoretic velocity times the Stokes friction)
acting along the path ${\caX}$.  Note
that the appropriate temperature $T_{\text{eff}}$ that
mediates between dissipation and entropy differs
from the local fluid temperature at the particle surface. Because of
the long-ranged hydrodynamic correlations, it has to be calculated as
the average of the temperature field emanating from the particle
weighted by the local dissipation, in the whole solvent
volume. General analytic expressions for $T_{\text{eff}}$ are provided
by the theory of non-isothermal Brownian motion \cite{fal14,fal14b}.

The effective temperature also quantifies the
  trade-off between the dissipation due to propulsion, $Q \equiv \mean{s}T_{\rm
    eff}= \mean{j_\|}v_{\rm p} \zeta t$, and the squared
  relative uncertainty in the particle current, $\epsilon^2 \equiv
  (\langle j_\|^2 \rangle- \mean{j_\|}^2)/\mean{j_\|}^2 \simeq 2
  D/(v_{\rm p} \mean{j_\|}t)$, namely
\begin{align}
\epsilon^2 Q \simeq 2 T_{\rm eff}.
\end{align}
This (saturated) thermodynamic uncertainty relation \cite{bar15} follows from
Eq.~(\ref{ft}) and the fact that $P(J_\|)$ is found to be well approximated by a Gaussian.
\begin{figure}
\centering
\includegraphics[width=\columnwidth]{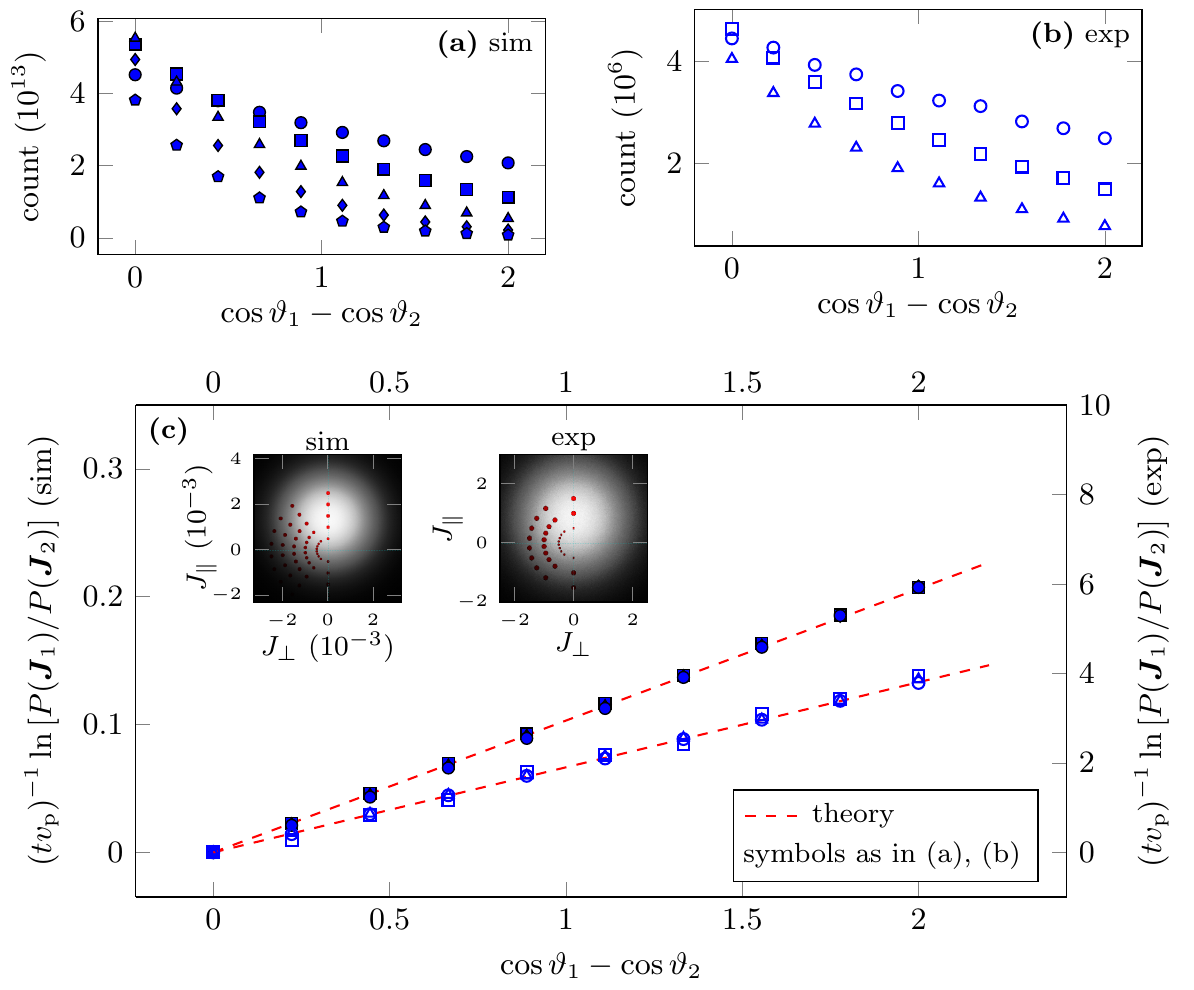}
\caption{\emph{Upper panel:} bin-counts for particle current,
    Eq.~(\ref{eq:current}), for various $J=|\vec J|$, from simulation
    (a) and experiment (b). \emph{Lower panel:}  test of the vectorial
    fluctuation theorem, Eq.~(\ref{logratios}), 
    using these bin-counts as proxies for $P(\vec J)$. 
    \emph{Insets:} heatmaps for $\vec J$, 
    red dots representing the bins from (a), (b).}
\label{fig:spatialFR}
\end{figure}

We finally turn to the validation of the spatial fluctuation theorem,
Eq.~\eqref{fts}.  In a co-rotating frame,
Eq.~\eqref{Langevin1} reads $\dot{\vec r}'= v_{\text{p}}
\vec n' + \sqrt{2D} \vec \xi_t$ with $\dot{ \vec r}' = \vec R \cdot
\dot{\vec r}$ and $\vec R(t)$ a time-dependent rotation matrix 
defined such that $\vec n' \equiv \vec R \cdot \vec n$ is a constant
versor arbitrarily chosen as the initial particle orientation.
Self-propulsion now shows up as the constant vector $v_\text{p}\vec
n'$. Without it, the particle would simply perform isotropic diffusion,
and the breaking of this spatial symmetry
gives rise to the spatial fluctuation relation \eqref{fts}, as much as
the breakdown of time-reversibility gives rise to the standard
fluctuation theorem \eqref{ft}.

To show this, we follow the procedure leading to Eq.~(\ref{logratio}), 
and consider the weights 
\begin{equation*}
P[\vec M{\caX'}] \propto \exp{\left(-\frac{1}{4D} \int_0^t \diff \tau
  \,(\dot{\vec r}'(\tau)-v_{\text{p}} \vec M \cdot \vec n' )^2\right) } ,
\end{equation*}  
for paths $\vec M{\caX'}$ that only differ in the particle orientation
$\vec M \cdot \vec n'$, where $\vec M$ is a constant rotation matrix
conserving the norm $(\vec M\cdot \vec n')^2 = {\vec n'}^2=1$. 
They are related by
\begin{equation}\label{ratio_ref}
  P[{\caX'}] = 
  P[{\vec M \caX'}]\exp\left(\frac{ v_{\text{p}}t}{2 D}
  {\vec n'} \cdot (\vec 1-\vec M^{-1}) \cdot \vec j [{\caX'}]\right),
\end{equation}
where $P[\caX']=P[{\vec M \caX'}]\big|_{\vec M= \vec 1}$, and
\begin{equation}\label{eq:current}
 \vec j[{\caX'}] \equiv \frac 1 t \int_0^t \diff \tau \,\dot{\vec r}'(\tau),
 \end{equation}
is the particle current relative to its instantaneous orientation $\vec n(t)$. 
After multiplying Eq.~\eqref{ratio_ref} by $\delta (\vec j[\caX']-\vec
J_1)$ and summing over trajectories, some algebra yields
\begin{equation}\label{logratios}
\frac{1}{t}\ln \frac{P(\vec J_1) }{P(\vec J_2)}= {\frac{ v_{\text{p}}}{2D} J (\cos \vartheta_1- \cos \vartheta_2) }\;.
\end{equation}
This spatial fluctuation relation expresses an exact symmetry between
the probabilities to observe currents $\vec J_i$ of equal
magnitude $J$ in different directions specified by their angles
$\vartheta_i$ with the versor $\vec n'$. Its equivalence with
Eq.~\eqref{fts} follows from $\vec J_i \cdot \vec n'= J\cos
\vartheta_i$ by identifying $\vec \epsilon \equiv v_{\text{p}}
\vec n'\zeta/(2 T_{\text{eff}})$, where one again recognizes the
dissipative non-isothermal driving. Again, it is valid for all times
$t$, provided that the trajectory is sampled on the diffusive time
scale. And it contains the scalar fluctuation relation,
Eq.~\eqref{logratio}, as the special case $\vartheta_1=0$,
$\vartheta_2=\pi$, i.e., $\vec J_1 = -\vec J_2=J_\| \vec
n$. Figure~\ref{fig:spatialFR}c shows that it is in excellent
agreement with our MD simulations and experiments.

In summary, we have verified the validity of scalar and vectorial
fluctuation relations for a self-propelled colloidal particle
suspended in a nonequilibrium solvent. This extends related recent
work \cite{kum15}, which could not conclusively settle the issue for
the case of an externally driven granular particle.  Using a minimal
Markovian model, we could recast our results in an intuitive form,
revealing that the breaking of the underlying microscopic space-time
symmetry is precisely quantified by the entropy production due to
swimming. The latter may be written as the energy dissipated to a
fictitious equilibrium bath at an effective temperature predicted by
the theory of hot Brownian motion.  The robustness of the established
fluctuation relations against some stochasticity in the driving and
the long-term memory and nonequilibrium character of the solvent
fluctuations suggests that the assumptions evoked by standard
derivations of fluctuation theorems are sufficient, but may actually
not all be critical for their successful application.

We acknowledge funding by the Deutsche Forschungsgemeinschaft (DFG). G.~F. thanks M.~Polettini and C.~P\'{e}rez-Espigares for discussions.

\bibliography{JanusFR}
 
\end{document}